\documentclass[12pt,a4paper]{article}

\usepackage[utf8x]{inputenc}
\usepackage[T1,T2A]{fontenc}
\usepackage[russian,english]{babel}

\usepackage[a4paper,top=2.5cm,bottom=2.5cm,left=2.5cm,right=2.5cm,marginparwidth=1.75cm]{geometry}

\usepackage{amssymb,amsmath}
\usepackage{slashed}
\usepackage{mathtools}
\usepackage{feynmf}

\usepackage{mathrsfs}
\usepackage{graphicx}
\usepackage{simpler-wick}

\usepackage[export]{adjustbox}
\usepackage[colorinlistoftodos]{todonotes}
\usepackage[colorlinks=true, allcolors=blue]{hyperref}
\usepackage{hyperref}
\usepackage{authblk}

\usepackage{tikz}
\usepackage{subfigure}
\usepackage{indentfirst} 
\usepackage{cite}
\usepackage{gensymb}

\graphicspath{{Pics/}}

\title{Neutrino Cooling of Primordial Hot Regions}

\author[*]{K.M. Belotsky}
\author[ ]{S.G. Rubin}
\author[ ]{M.M. Elkasemy}
\affil[ ]{National Research Nuclear University MEPhI (Moscow Engineering Physics Institute), 115409, Kashirskoe shosse 31, Moscow, Russia}
\affil[*]{\small E-mail: k-belotsky@yandex.ru}
\date{}

\begin{document}

\maketitle

\abstract{
The effect of neutrino cooling of possible primary regions filled by the hot matter is discussed. Such regions could contain the primordial density inhomogeneities of different origin and survive up to modern epoch. We show that the final temperature of such region is $\sim 10\, {\rm keV}$ provided that the initial temperature is within the interval $10\, {\rm keV} ... 100\, {\rm MeV}$. The cooling is realized due to the nuclear reactions containing $n-p$ transition. The lower limit $10\, {\rm keV}$ is accounted for by suppression of the reactions rates because of threshold effect and $n$, $e^{\pm}$ density diminishment. 
}

\large
\section{Introduction}
The physics of the early Universe is one of the main tool for study of both the elementary particle physics and the cosmology. Any imprints of the physics at early stage and phenomena which are responsible for them, deserves careful analysis. 

In this paper, we pay attention to the local heated areas in the early Universe. There are some observations \cite{alihaimoud2019electromagnetic} favouring their existence. 
Similar objects have been hypothetically discussed earlier \cite{dubrovich2012cosmological, kumar2019cmb, kogut2019cmb}. They could be separated from the Hubble flow if their density is high enough. For instance, such regions could be formed by clusters of primordial black holes (PBH) \cite{Belotsky_2019, alihaimoud2019electromagnetic}. In the following, the origin of the regions is not specified.

We assume that the matter including the dark matter was captured by the gravitational 
forces before the star formation, when the temperature of the Universe media was high enough. Such regions are supposed to be of finite size being teared off the hot plasma.  There are several reasons for an unusual temperature evolution within such regions 
 which we discuss below. 
 
Because of its finite size, the region can be transparent for neutrinos even at $T_0 > 1$ MeV while the Universe does not.

Essential thing is the following. Gravity of the region can hardly keep electrons and positrons at the
temperature $T_0\lesssim 1$ MeV. However, diffusive character of their propagation as well as of photons will effectively prevent 
their escape \cite{dubrovich2012cosmological}. The escape time of electrons from the region of the size 
$R$ with temperature $T\sim m_e$ 
would be \cite{lifshitz2006physical}
$$t\sim \frac{R^2}{D}\sim R^2T\alpha^2\sim 10^{24}\, {\rm year}. $$
For numerical estimate we supposed that $R\sim 1$ pc as has been obtained at some model parameters for PBH cluster \cite{Belotsky_2019}, electron velocity $v$ and Coulomb logarithm $L$ were supposed to be of order 1, diffusion coefficient was estimated from electron travelling length \cite{lifshitz2006physical} $D\sim \lambda v\sim T^2/\alpha^2n$ with $n\sim T^3$ being electron number density. As one can see that even for $R\ll 1$ pc diffusion time is much bigger than the Universe lifetime. 

In further, we will refer to $T_0$ as initial temperature. Initial mass of the region can be as big as the mass of matter inside the horizon at the temperature $T_h>T_0$ when the region started to form, i.e. roughly $M\sim 10^6 ({\rm MeV}/T_h)^2M_{\odot}$. 
Therefore $T_h>10...100$ MeV is less interesting because of small area mass, though the considered effects here should be also applicable to this case.

After virialzation the regions can be heated or cooled due to different processes inside them. Among then are neutrino cooling, nuclear reactions, radiation of the hot plasma and the stars formed inside the region \cite{Dolgov_2017}, gravitational dynamics of the system, shock waves, diffusion of matter, variation of the vacuum state while the region is born  \cite{Belotsky:2017txw}, 
energy transfer from collapsing walls \cite{BEREZIN198391,khlopov1998formation,rubin2000primordial,Deng_2018}, accretion, Hawking evaporation. The last mechanisms imply existence of PBH  \cite{dolgov1993baryon,Dolgov_2018}.
 
Neutrino cooling could be the most effective mechanism among the listed ones in some energy range.

Neutrino cooling is realized due to reactions of weak $p\leftrightarrow n$ transitions. The reaction rate is slowing down when the area is cooling which is the result of the threshold effects and decreasing of $n$ and $e^{\pm}$ densities. Here we show that the temperature tends to some fixed magnitude below which neutrino cooling is ineffective.

\section{Reactions inside heated area}

The size, mass and temperature of a volume separated from the cosmological expansion depend on many circumstances. When choosing the initial conditions, we will partially follow the paper \cite{Belotsky_2019} where the mass of the  initially trapped matter is varied in wide range, $10^4-10^8 M_{\odot}$. 
Initial parameters of the hot area basically used here are: the size of the region is about 1 pс, its mass $10^4\,M_{\odot}$, initial temperature lies in the interval  $T_0\sim 1\text{keV}\div 10 \text{MeV}$ (at higher temperature 
cluster can start to absorb neutrino). The change of the parameters in wide range does not affect the qualitative effect.

Consider the cooling of the heated area due to the neutrino radiation. The basic reactions of neutrino production are supposed to be the following
\begin{equation}
e^- + p \rightarrow n + \nu_e,
\label{ep}
\end{equation}
\begin{equation}
e^+ + n \rightarrow p + \bar\nu_e,
\label{en}
\end{equation}
\begin{equation}
e^+ + e^- \rightarrow \nu_{e,\mu,\tau}+\bar{\nu}_{e,\mu,\tau},
\label{ee}
\end{equation}
\begin{equation}
n \rightarrow p+e^- + \bar\nu_e. \label{n}
\end{equation}

The necessary formulas which we used in calculations of the rates of given reactions are as follows. The neutrino production rate per unit volume, $\gamma_i\equiv \Gamma_i/V$, for each reaction are respectively
\begin{eqnarray}
\label{gamma}
\gamma_{ep}=n_{e^-}n_p\sigma_{ep}v,\\
\gamma_{en}=n_{e^+}n_n\sigma_{en}v,\\
\gamma_{ee}=n_{e^-}n_{e^+}\sigma_{ee}v,\\
\gamma_n=\frac{n_n}{\tau_n}.
\end{eqnarray}
Here $n_i$ is the number density of the respective species, $\sigma_{ij}$ is the cross section of interacting particles $i\, {\rm and }\,j$, $v$ is their relative velocity, which are supposed to be equal 1 for all the reactions, and $\tau_n\approx 1000$ sec is the neutron lifetime.

The number densities are approximately described by the following formulas
\begin{eqnarray}
n_p=\frac{n_B}{1+\exp\left(-\frac{\Delta m}{T}\right)},\;\;\;\;\; n_n=n_p(T)\exp\left(-\frac{\Delta m}{T}\right),\\
n_{e^-}=n_e^{eq}(T)+\Delta n_e,\;\;\;\;\;n_{e^+}= n_e^{eq}(T) \exp\left(-\frac{m_e}{T}\right),\\
n_B\equiv n_p+n_n=g_B\, \eta n_{\gamma}(T_0),\;\;\;\;\; \Delta n_e\equiv n_{e^-}-n_{e^+}=n_p.
\label{nd}
\end{eqnarray}
Here  
$\eta=n_B/n_{\gamma}\approx 0.6\cdot 10^{-9}$ is the baryon to photon relation in the modern Universe, $g_B\sim 1$ is the correction factors of that relation due to entropy re-distribution, $n_{\gamma}(T)=\frac{2\zeta(3)}{\pi^2}T^3$ and $n_e^{eq}=\frac{3\zeta(3)}{2\pi^2}T^3\approx 9\cdot 10^{25}\left(\frac{T}{100 {\rm keV}}\right)$ are the equilibrium photon and electron number densities respectively,  $\Delta m=m_n-m_p=1.2$ MeV. One should pay attention that $n_{\gamma}$ depends on $T_0$ rather than $T$.

The following approximate formulas
\begin{eqnarray}
\sigma_{en}=\sigma_{ee}=\sigma_w,\\
\sigma_{ep}=\sigma_w\, \exp\left(-\frac{Q}{T}\right),
\label{sigma}
\end{eqnarray}
 are used for the cross sections. Here
$\sigma_w \sim G_{\rm F}^2 T^2$, $Q=m_n-(m_e+m_p)=0.77$ MeV effectively takes, through exponent, into account threshold effect in respective reaction, $G_{\rm F}=1.166\cdot 10^{-5}\,\rm{ GeV}^{-2}$ is the Fermi constant.

\hspace{5 mm}

The backward reactions for \eqref{ep}-\eqref{n} are suppressed since neutrino freely escape the cluster. This is essential feature of the considered case distinguishing it from the processes in the early Universe. Indeed, the neutrino would scatter off $e^{\pm}$, with the mean free path  $\lambda_{\nu}=1/n_e \sigma_{\nu e} \simeq 10^{24}$ cm at the temperature $T= 100$ keV.  Hence, $\lambda_{\nu}$ of the neutrino inside the cluster would be much larger than its size, 1 pc in our case.  
This justifies the importance of study the energy outflow by neutrinos and subsequent cooling of the media inside the cluster.

Note one more that the densities of the relativistic matter components like $e^{\pm}$ and $\gamma$ are functions of the current temperature $T$ (which is varied due to neutrino cooling and possible other effects), contrarily to the baryon density. The latter is defined by initial temperature $T_0$ and almost unchanged with time.

\section{Temperature evolution}

The temperature balance
\begin{equation}\label{eq5}
\Delta Q = \delta U
\end{equation}
follows from the first law of thermodynamics.
Here $\Delta Q$ and $\delta U$ are the heat outflow due to neutrinos and the 
diminution of inner energy of the matter inside the cluster respectively. Applying it to our case, we obtain the energy conservation in the form 
\begin{equation}\label{temp1}
-(\gamma_{en} + \gamma_{ep} + \gamma_{ee} + \gamma_{n} )  E_{\nu} dt = 4b T^3 dT,
\end{equation}
where $E_{\nu}$ is the energy of outgoing neutrino which is supposed to be $\sim T$, 
$b = 
\pi^2 /15$ is the radiation constant.

Integrating Eq\eqref{temp1}, one can get time dependence of the temperature in an implicit form
\begin{equation}\label{temp3}
\Delta t = -4 b \int_{T_0}^T  \frac{T'^2dT'}{\gamma_{en} + \gamma_{ep} + \gamma_{ee} + \gamma_{n}}.
\end{equation}
Here explicit dependence of the functions $\gamma_i$ on $T$ is the following: $\gamma_{en}=C_1\cdot T^5\exp\left(-\frac{Q}{T}\right)$, $\gamma_{ep}=C_2\cdot T^5\exp\left(-\frac{m_e+\Delta m}{T}\right)$, $\gamma_{ee}=C_3\cdot T^8\exp\left(-\frac{m_e}{T}\right)$, $\gamma_{n}=C_4\cdot \exp\left(-\frac{\Delta m}{T}\right)$, where exact view of $C_i$ follows from Eqs.\eqref{gamma}--\eqref{sigma} and their specifications in the text. The coefficients $C_{1,2,4}$ also contain the multiplier $\left[1+\exp\left(-\frac{\Delta m}{T}\right)\right]^{-1}$.

Numerical solutions of Eq.\eqref{temp3} are represented in Fig.\ 1 left, where different curves correspond to different initial temperatures $T_0$. 
As seen,
the temperature falls sharply around $T\sim 100$ keV and then smoothly subsides to $\sim 10$ keV. We call the time when this temperature is reached as a cooling time $t_{\rm cooling}$. It is drawn in
figure 1 right.

\begin{figure}[ht]
    \subfigure{
    \includegraphics[width=0.49\textwidth]{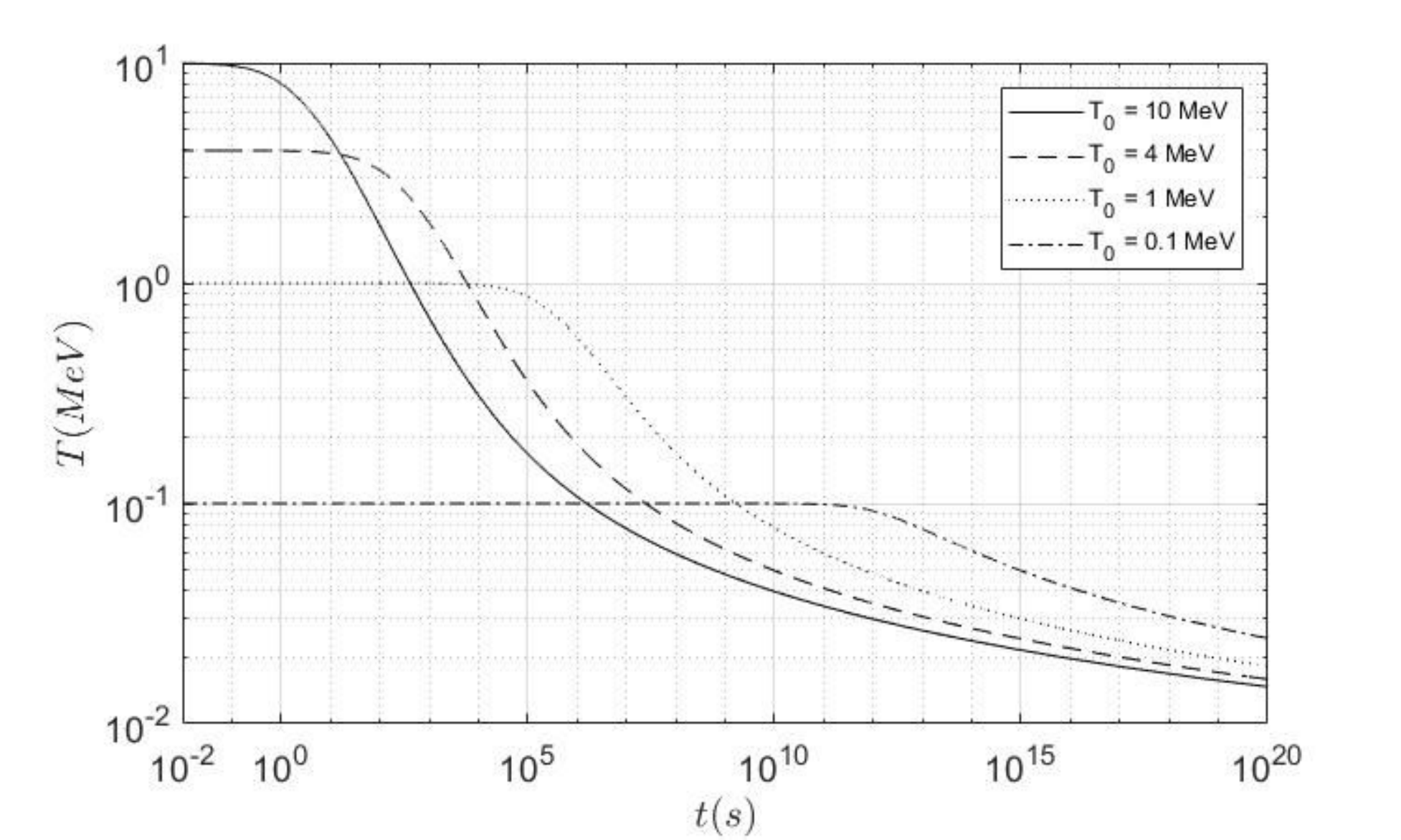}}
    \subfigure{
    \includegraphics[width=0.49\textwidth]{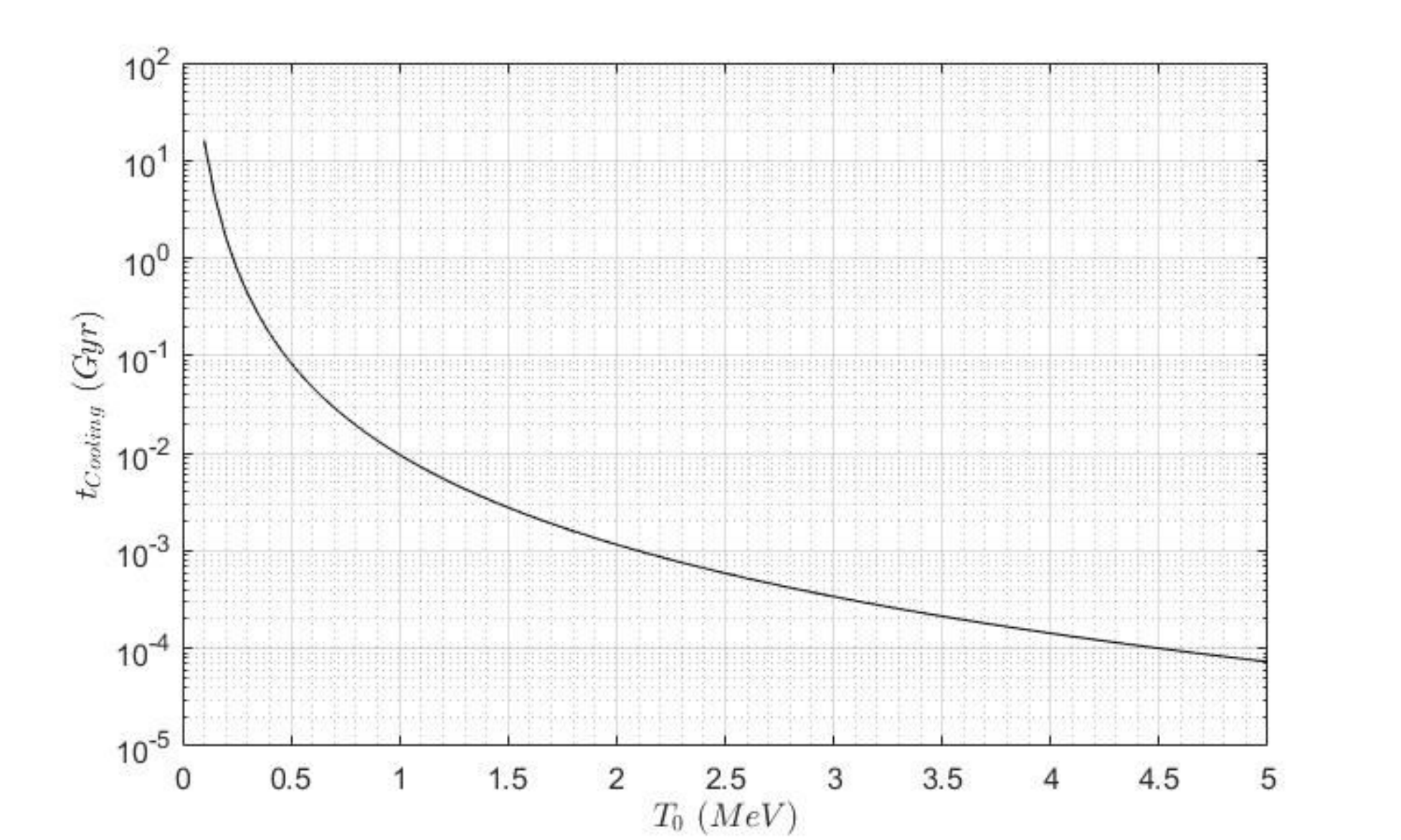}}
    \caption{Left: The time behaviour of the temperature inside the heated area. Right: Cooling time $t_{\rm cooling}$ of media inside the heated area depending on the initial temperature $T_0$.} 
    \label{fig}
\end{figure}

As one can see, neutrino cooling effect can be essential for the heated areas with initial temperature $T_0\gtrsim 100$ keV.   

Also, the line in the Fig.1 for the initial temperature $T_0=0.1$ MeV virtually does not change with time, unlike the curves for higher $T_0$ even when they go below 0.1 MeV. The reason is that at the higher initial temperature we have the higher initial densities of all matter components. At the same time, density of baryons does not change and $\Delta n_e$ changes slowly in time in the considered gravitationally separated area. Thus reactions proceed faster for higher $T_0$ even when $T$ becomes lower 0.1 MeV.

\section{Conclusion}
The relic hot regions are widely discussed in the literature, some of them are mentioned in the Introduction. Their existence could open new ways for research and study observational effects. 

In this letter, we study the neutrino mechanism of cooling of such regions. We have shown that these regions are cooling down to the value $\sim 0.01\div 0.1$MeV due to neutrino emission. In the following, the temperature varies slowly up to the present times.

There are other heating and cooling effects that seem less decisive than the neutrino emission process. Nevertheless, their study should be carried out in the future.

It is worth paying attention to PBH as a possible origin of such regions. This is unique phenomena indicating possible deviation from the standard astrophysical scenario (see, e.g., recent works \cite{carr2020constraints,hasinger2020illuminating,hawkins2020signature,garcia2018primordial,clesse2017seven}). They are often involved in attempts to unravel a variety of the cosmological puzzles.

PBH cluster could naturally account for existence of such regions \cite{Belotsky_2019, alihaimoud2019electromagnetic}. Moreover, the domain-wall mechanism of the cluster formation \cite{rubin2000primordial,rubin2001formation,dokuchaev2004quasars} can lead to additional heating of the matter inside it \cite{BEREZIN198391,khlopov1998formation,rubin2000primordial,Belotsky:2017txw}. 
Further analysis of observational effects of such regions, 
as well as point-like cosmic gamma-ray sources \cite{Belotsky_2011} and gravitation lensing events \cite{Toshchenko:2019mth}, would provide an important test of the hypothesis on the PBH cluster existence.

Here we do not concentrate on PBH origin of the considered regions. The neutrino cooling   is one of the process accompanying the PBH cluster formation. It deserves separate complex investigation.


\vspace{3 mm}
\textit{Acknowledgement}

Our special thanks to M.\ Khlopov for useful discussion.
We would like to thank A.\ Kirillov, V.\ Nikulin
for the interest and help in some issues.

The work was supported by the Ministry of Science and Higher Education of the Russian Federation
as part of the Program for Improving the Competitiveness of the MEPhI (project no.\ 02.a03.21.0005),
and also (work of K.M.B.) by project No 0723-2020-0040 ''Fundamental problems of cosmic rays and dark matter``.
The work of S.G.R.\ is supported by the RFBR grant 19-02-00930 and 
is performed according to the Russian Government Program of Competitive Growth of Kazan Federal University.
%

\bibliographystyle{JHEP}
\bibliography{mybibliography}
\end{document}